\documentclass[conference, a4paper]{IEEEtran}
\IEEEoverridecommandlockouts
\usepackage{cite}
\usepackage{amsmath,amssymb,amsfonts}
\usepackage{algorithmic}
\usepackage{graphicx}
\usepackage{textcomp}
\usepackage{xcolor}
\def\BibTeX{{\rm B\kern-.05em{\sc i\kern-.025em b}\kern-.08em
    T\kern-.1667em\lower.7ex\hbox{E}\kern-.125emX}}
    
\graphicspath{{figures/}}
\usepackage{dsfont}
\usepackage{psfrag}
\usepackage{pgfplots}
\begin{document}

\newcommand{\e}[1]{\bar{#1}}
\newcommand{\CE}[2]{\mathbb{E}\left[ #1 \ \middle| \ #2 \right]}
\newcommand{\E}[1]{\mathbb{E}\left[ #1 \right]}
\newcommand{\tr}[1]{\tilde{#1}}
\newcommand{\xp}[1]{\tilde{#1}}
\newcommand{\myfigurespace}{\vspace{0mm}}

\title{A linear algorithm for reliable predictive network control}

\author{\IEEEauthorblockN{Richard Schoeffauer and Gerhard Wunder}
\IEEEauthorblockA{\textit{Heisenberg CIT Group, Free University of Berlin; Berlin, Germany} \\
richard.schoeffauer@fu-berlin.de, g.wunder@fu-berlin.de}
}

\maketitle

\begin{abstract}
This paper introduces a novel control approach for network scheduling and routing that is predictive and reliable in its nature, yet builds upon a linear program, making it fast in execution. First, we describe the canonical system model and how we expand it to be able to predict the success of transmissions. Furthermore, we define a notion of reliability and then explain the algorithm. With extended simulations, we demonstrate the gains in performance over the well known MaxWeight policy.
\end{abstract}


\section{Introduction}

The fifth generation of mobile communication, 5G, aims at not only enabling communication between billions of people around the globe but also connecting billions of devices. In this context, many boundaries are currently tackled by research, such as increasing data rates, providing security and many more.

This paper is dedicated to enhance performance of networked devices through a \textit{predictive} scheduling and routing of data packets through the network. Specifically, the here presented network control policy enhances the performance of interconnected robust model predictive controllers (RMPCs). The policy does not only schedule and route data, but as a novel feature, also predicts their time of arrival at their corresponding destination. In other words it predicts the communication delays. Signaling these communication delays ahead of the arrival of the actual data to the corresponding RMPCs facilitates them to enhance their control performance, as was first shown in \cite{hahn2018}.

As a reference network control policy, we will use the well known MaxWeight or MaxPressure policy, (from now on written as MW), first introduced by \cite{tassiulas1992}. In the last two decades, MW and other network control strategies were investigated intensively, e.g. in \cite{meyn2007}. However, focus has yet always remained at lowering overall delay while maintaining the property of maximum throughput, which makes MW such a good policy in the first place \cite{kasparick2017}. And though prediction of has been successfully used to improve overall delay in broadcast scenarios \cite{zhou2007}, for the best of our knowledge, trying to predict individual packet delays is a novel idea.

Another kindred topic is the so-called delay-constrained routing and scheduling. While originating from area of wired communication \cite{frangioni2017}, it has also been investigated for the wireless case \cite{lee2009}. However, results are yet limited to rather mathematical statements with limited use for practice.

In this paper we represent a fast algorithm to schedule and route information through a network and at the same time provide forecasts of specific delay times in a \textit{reliable} fashion. We build on the ideas of \cite{schoeffauer2018} where we described a first approach to yield reliable delay forecasts. However the algorithm that was presented was computationally expensive (due to its quadratic nature) and had worse performance in the achieved delay times (caused by a strictly repetitive activation of links). The algorithm designed in this paper, will eliminate these shortcomings.

\section{System model}
\label{sec:system_model}

We make use of the standard queueing model, which is time discrete, integer valued and offers binary controls. This is the appropriate choice for packet level modeling. Each of the $n$ agents in the network may hold multiple (data-) packets at a given time slot $t$, which are to be transmitted to other agents. Those packets are lined up in so-called queues
$q^i, i = 1,\dots n$.
E.g. $q^i = 4$ represents $4$ packets, waiting in queue $i$ (located at agent $i$). The vector of all queues will be denoted as $q \in \mathbb{N}^n$. With this model, sending packets from one agent to another is represented by decreasing and increasing queues at the corresponding agents. De- and increasing can be done by a vector $r^j \in \{-1,0-1\}^{n}$, $j = 1, \dots m$, which is called a link; the matrix of all links is called the routing matrix $R \in \{-1,0-1\}^{n\times m}$. In each time slot, we can choose to activate a link through a binary control vector $u \in \{0,1\}^m$, so that the system evolves like
\begin{equation}
\label{eq::standart_system_evolution}
q_{t+1} = q_t + R_t u_t + a_t \ , \quad \text{s.t.} \quad  Cu_t \leq \mathds{1}  \ , \quad  q_{t+1} \geq 0
\end{equation}
where $\mathds{1}$ is a vector of ones with appropriate dimension.
The arrival $a_t \in \mathbb{N}^n$ expresses an influx of information to the system and is usually of stochastic nature with expectation $\E{a_t} = \e{a}$. The constituency matrix $C$ prohibits to activate all controls simultaneously, and naturally queues can only hold a positive number of packets, giving rise to the positivness constraint.
Note, that it is a key feature of \textit{wireless} links to have a (stochastically) time dependent \textit{current} routing matrix $R_t$. In each time slot, $R_t$ is a random selection of the columns from a known routing matrix $R$, where non selected columns are set to $0$ in $R_t$. This can be expressed via a diagonal probability matrix $M = \operatorname{diag}_{i=1 \dots m} \{p^i\}$, holding the transmission success probabilities $p^i \in [0,1]$ for each link. Performing a Bernoulli trial $\mathbb{B}[]$ on $M$ and multiplying it to $R$ gives the mentioned selection of links: $R_t = R \cdot \mathbb{B} \left[ M \right]$. Notice, that a controller only knows $R$ and $M$ but not $\mathbb{B}[M]$, thus not every activation (scheduled  transmission) will succeed. 
The problem lies in finding the best suited control to steer the data to its destination, though not fully knowing the outcome of the Bernoulli trial.

In order to use meaningful predictions of future behavior, we enhance this standard model by defining a whole set of probability matrices $M_i \in \mathcal{M}$ instead of only one (as already done in \cite{schoeffauer2018}). In each time slot, the system uses one $M_i$, according to a discrete time markov chain, that evolves on the index set $\mathcal{I}(\mathcal{M})$ as $\sigma_t = \mathbb{M} \left( \mathcal{I}(\mathcal{M}) , P , \sigma_0 \right)$, $P$ being the transition matrix and $\sigma_0$ the initial markov state. Hence we get $R_t = R \cdot \mathbb{B} \left[ M_{\sigma_t} \right]$. We assume that the controller has full knowledge of $\mathcal{M}$, $P$, and $\sigma_t$, i.e. knows the expected transmission success probabilities of all links and all future times. Of course, the network controller may be required to measure these parameters before being able to control the network. In this transient state, the network may be controlled be the MW policy.

\section{Reliable predictive network control}

To predict packet transmissions in the system model, the network controller internally uses a slightly different model to predict the flow of the packets. This so called \textit{prediction model} is shown in Fig. \ref{fig::waterfall} and will be described in this section. Two major circumstances give rise to the prediction model:

\begin{figure}[htbp]
\psfrag{rw}[c][c]{actual setting}
\psfrag{cm}[c][c]{system model}
\psfrag{pm}[c][c]{prediction model}
\psfrag{f11}[c][c]{$p_1^1$}
\psfrag{dp}[lt][lt]{\shortstack{data\\packet}}
\psfrag{gg}[rt][rt]{$\omega_1^1$}
\includegraphics[width=0.95\linewidth]{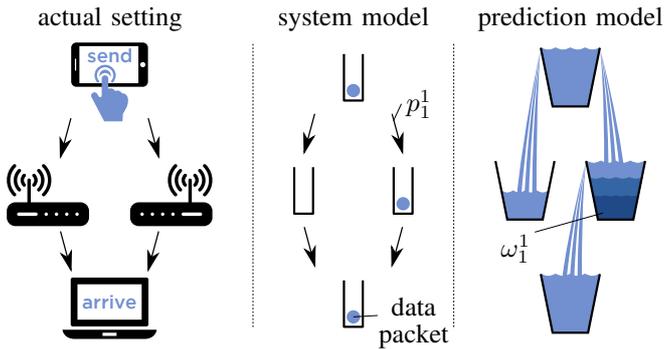}
\vspace*{2mm}
\caption{The different levels of abstraction}
\label{fig::waterfall}
\end{figure}

1) Two packets, simultaneously residing at the same agent, may still be intended for different destination agents. Hence, in order distinguish different packets mathematically, each one of them has be cast with its own copy of the system model~\eqref{eq::standart_system_evolution}. These copies will be called subsystems. Suppose that $s$ is the number of subsystems currently in use (i.e. $s$ packets are currently present in the network), then with slight abuse of notation we can define stacked versions of all quantities and write the evolution of the prediction model as:
\begin{equation}
\label{eq::system_evo}
q_{t+1} = q_t + R_t u_t
\end{equation}
where from now on
\begin{equation}
\begin{gathered}
R_t := I_s \otimes \big[ R \cdot \mathbb{B} \left[ M_{\sigma_t} \right] \big]
\\
u_t^T := \left( {u^{(1)}_t}^T , \dots  {u^{(s)}_t}^T \right) 
\ , \ 
q_t^T := \left( {q^{(1)}_t}^T , \dots  {q^{(s)}_t}^T \right)
\end{gathered}
\end{equation}
and $(\cdot)^{(i)}$ corresponds to the $i$-th subsystem; $I$ denotes the identity matrix and from now on we use $n$ and $m$ to denote the dimensions of the stacked vectors. Note that $C$ may change in a different way depending on the scenario.
Here, we can ignore the arrival $a_t$ since any new packet arriving at the system will immediately lead to another subsystem being cast and stacked on top the current prediction model. E.g. if agent $i$ signals the initialization of a new packet, then subsystem $(s+1)$ will be cast and the packet will be represented by initializing $q_t^{(s+1)}$ with a $1$ at the $i$-th component. In the same way, a subsystem can be erased from the stack, once the agent, the packet was intended for, signals (to the controller) its successful delivery.

2) The agents, here RMPCs, are only able to use the forecasts of packet-specific communication delays, if these forecasts are reliable (so that the RMPCs can still guarantee robustness). Therefore, our policy will predict the flow through the network with explicit consideration of possible transmission failures. Specifically, in its prediction for future activations, the algorithm will repeat activating a link, until the accumulative transmission failure probability falls beneath a \textit{single-transmission failure-probability threshold} $\tau$.
In other words, if 
$f_t^i = 1 - \mathbb{E} \left[ p_t^i \middle| \sigma_0 \right] $ 
are the expected failure probabilities of link~$i$, then we reliably transmitted a packet over that link (in the prediction), only if
\begin{equation}
\label{eq::rel_constraints}
\prod_{t = 0}^{H-1} {f_t^i}^{u^i_t} \leq \tau
\qquad
\Longrightarrow
\qquad
\sum_{t=0}^{H-1} u^i_t \log_{\tau} f_t^i \geq 1
\end{equation}
where $u^i_t$ is the corresponding control variable for this link and $H$ is the prediction horizon. It is an easy task, to derive the expected failure probabilities from the discrete time markov chain via
\begin{equation}
\begin{pmatrix}
f^1 \\ & \ddots \\ & & f^m
\end{pmatrix}_t 
=
I_m -
\left[ \sigma_0 P^t \otimes I_m \right]
\begin{pmatrix}
M_1
\\
\vdots
\\
M_p
\end{pmatrix}
\end{equation}

Now we return to the network controller. It is implemented as an MPC, meaning that in \textit{each} time slot, the controller minimizes a cost function (influenced by the prediction model) to yield a control trajectory $\tr{u}^T = \left( u_0^T ,\dots u_{H-1}^T \right)$ over a horizon $H$ but then only applies the first component ($u_0$) to the system. Here, the current time step is set to $0$ for ease of notation.
Usually, MPC objective functions are quadratic in nature, leading to semi definite programs. Since network control has to happen very fast (depending on the granularity of the data), a main contribution in this paper is to devise an algorithm that is specifically designed to be a binary linear program which is solvable in polynomial time \cite{munapo2016}.

The intuition behind the algorithm equals a waterfall, always filling the queues in direct vicinity of already filled ones. As a first step we introduce the reliability $\eqref{eq::rel_constraints}$ as a constraint. Let
\begin{equation}
\omega_t^i = \max \left \{ 
\ \log_{\tau} f_t^i  \ \  | \ \ 1 \
\right \}
\end{equation}
then we can define
\begin{equation}
\Omega_C = \begin{pmatrix}
\operatorname{diag}_i \left\{ \omega^i_0 \right\}
& \dots &
\operatorname{diag}_i \left\{ \omega^i_{H-1} \right\}
\end{pmatrix}
\in \mathbb{R}^{m \times mH}
\end{equation}
Forcing $\Omega_C \tr{u} \geq 1$ will make all $\tr{u}$ guarantee reliable activations and hence reliable forecasts as described earlier. However, applying this to our system evolution so far could contradict the positiveness constraint on $q$. E.g. having three scheduled activations $u^i$ of the link $i$ (in order to be reliable) would result in a negative queue state of $1-3=-2$ at the link-origin queue and suggest the presence of $0+3=3$ data packets at the link-destination queue. 
To compensate, we bring two changes to the prediction model.
First, we change the $r^i$ (links) to not \textit{de}crease any queue states at all. Second, we multiply the $r^i$ with their corresponding $\omega_t^i$ values, resulting in the queue vector being real valued ($q_t \in \mathbb{R}^n$). Incorporating these changes into \eqref{eq::system_evo} while simultaneously considering the evolution for the whole prediction horizon yields
\begin{equation}
\begin{pmatrix}
q_1 \\ \vdots \\ q_H
\end{pmatrix}
= \begin{pmatrix}
q_0 \\ \vdots \\ q_0
\end{pmatrix}
+
\begin{pmatrix}
R^+ \\
\vdots & \ddots \\
R^+ & \dots & R^+
\end{pmatrix}
\Omega_E \tr{u}
\end{equation}
with the two definitions
\begin{equation}
R^+_{(i,j)} = \max \{ 0 , R_{(i,j)} \}
\end{equation}
and
\begin{equation}
\Omega_E = \operatorname{diag}_{t = 0 \dots H-1} \left\{ \operatorname{diag}_{i=1\dots m} \left \{ \omega^i_t \right \} \vphantom{\frac{1}{2}} \right \}
\end{equation}

Note, that ignoring to decrease queues is only viable, because we are using separate subsystems for each single packet and therefore need only to consider the propagation of the packet but not what happens to queues that have already been passed by it. This way, we also avoid further constraints for the positiveness of the queues.
Furthermore, using $\omega_t^i$ as weights means that a a packet is predicted to have successfully been transmitted over a link, if the link-destination queue is filled exactly to or beyond $1$.
However we are still missing two main ingredients for the prediction model to work: on the on hand, queues filled just beyond $1$ are not supposed to be filled any further. On the other, only those links can be activated, whose link-origin queue has been filled exactly to or beyond~$1$.

By the virtue of $u_t^i$ being binary, we can formulate both constraints in a linear manner. Let $T^d \in \{0,1\}^{m \times n}$ be a simple transformation matrix, that rearranges $q_t$ in such a way, that the $i$-th entry in $T^d q_t$ is the link-destination queue of link $i$, and define $T^o$ in the same way for the link-origin queues.
Then we get for the first constraint
\begin{equation}
\begin{aligned}
u_t &\leq 2 - T^d q_t \\ &= 2 - T^d q_0 - T^d R^+ \left( \Omega_E^0 u_0 + \dots + \Omega_E^{t-1} u_{t-1} \right)
\end{aligned}
\end{equation}
and for the second
\begin{equation}
\begin{aligned}
u_t &\leq T^o q_t \\ &= T^o q_0 + T^o R^+ \left( \Omega_E^0 u_0 + \dots + \Omega_E^{t-1} u_{t-1} \right)
\end{aligned}
\end{equation}
where $\Omega_E^t$ describes the part of $\Omega_E^t$ that corresponds to time slot $t$. Defining a block triangular matrix as
\begin{equation}
\Delta \left( T^dR^+ \right)
=
\begin{pmatrix}
0 \\
T^d R^+ & \ddots \\
\vdots & \ddots &  \ddots \\
T^d R^+ & \dots & T^d R^+ & 0
\end{pmatrix} 
\end{equation}
we can write this over the prediction horizon to yield
\begin{equation}
\left[ I_{Hm} + 
\Delta\left( T^dR^+ \right)
\Omega_E
\right]
\tr{u}
\leq
\mathds{1}_H \otimes \left[ 2 - T^d q_0 \right] 
\end{equation}
for the first and
\begin{equation}
\left[ I_{Hm} - 
\Delta\left( T^o R^+ \right)
\Omega_E
\right]
\tr{u}
\leq
\mathds{1}_H \otimes \left[ T^o q_0 \right]
\end{equation}
for the second constraint.
Together with the reliability constraint $\Omega_C \tr{u} \geq \mathds{1}$ and a suitable constituency constraint $\tr{C} \tr{u} \leq \mathds{1}$ this completes evolution and constraints of the prediction model. (In the end of this section, we will discuss the case, in which the constraints can not be fulfilled.)

This leaves us with the definition of a suitable objective function $J$. In a linear fashion, we use the weight vector ${\gamma \in \mathbb{R}^n, \gamma < 0}$ to assign rewards to filling any queue. The ``closer'' such a queue is to the subsystem-destination queue, the higher the reward it grants. With proper $\gamma$, the algorithm thus will automatically push the packet in the right direction. For simple networks, $\gamma$ can be constructed by hand. How to arrive at an optimal $\gamma$ is however still subject to research.
In any case, the objective function becomes
\begin{equation}
\label{eq::obj_function}
J = \sum_{t=1}^H \gamma^T q_t = \left[ \mathds{1}^T_H \otimes \gamma \right] 
\begin{pmatrix}
R^+ \\
\vdots & \ddots \\
R^+ & \dots & R^+
\end{pmatrix}
\Omega_E \tr{u}
\end{equation}

To summarize, the control policy consists of solving the following minimization problem in \textit{each} time step, while only applying the first component of the optimal control trajectory $\tr{u}$ (that minimizes the objective): 
\begin{equation}
\begin{gathered}
\min_{\tr{u}} \ \eqref{eq::obj_function}
%\left[ \mathds{1}^T_H \otimes \gamma \right] 
%\begin{pmatrix}
%R^+ \\
%\vdots & \ddots \\
%R^+ & \dots & R^+
%\end{pmatrix}
%\Omega_E \tr{u}
\\
\text{s.t.}
\\
\tilde{C} \tr{u} \leq \mathds{1}
\\
-\Omega_C \tr{u} \leq -\mathds{1}
\\
\left[ I_{Hm} - 
\Delta\left( T^o R^+ \right)
\Omega_E
\right]
\tr{u}
\leq
\mathds{1}_H \otimes \left[ T^o q_0 \right]
\\
\left[ I_{Hm} + 
\Delta\left( T^dR^+ \right)
\Omega_E
\right]
\tr{u}
\leq
\mathds{1}_H \otimes \left[ 2 - T^d q_0 \right] 
\end{gathered}
\end{equation}
Note that this is a binary linear program with linear constraints. Furthermore, any matrices can be precomputed offline, making it feasible to solve. Given an optimal $\tr{u}$ it is an easy task, to derive at the prediction of when packets will arrive at their corresponding destination.

For completeness we finally address some technicalities left open:
(1) There are cases in which the reliability constraint can not be fulfilled by any $\tr{u}$ at all (e.g. if $H$ is too small). As a solution, we append a dummy control $u_D$ to $\tr{u}$, which has no influence on the prediction model evolution and is penalized with suitable weights in $J$. Writing the reliability constraint as
$ \left[ \tr{u}^T \ | \ u_D^T \right] \left[ \Omega_C \ | \ I_m \right]^T \geq \mathds{1}^T$ guarantees a feasible solution.

(2) Once the subsystem-destination queue $q^*$ has been filled, no further activations in this subsystem are to be scheduled. To this end, we engineer a dummy queue $q^D$ and a link from $q^*$ to $q^D$. We reward filling of $q^D$ highly in $J$ and disable the constraint so that it can be filled without limit. Making activation of the dummy link disjunct to any other activation in the subsystem will result in the desired behavior.

(3) To ease the understanding we omitted a constraint that would force the policy to yield only reduced or equal delay times at consecutive time slots. This constraint has to be added in order for the policy to stay consistent with its forecasts.

(4) As a general way of defining $\gamma$, one can use the Dijkstra algorithm on each subsystem. Doing this, the weights of the links should be defined as the number of consecutive repetitions necessary to fulfill reliability over that link. Here, one can work with time-averaged transmission success probabilities. Offsetting the derived shortest paths for each queue then yields the reward coefficients for $\gamma$.

(5) For the entire algorithm to work, we assume that all agents store their received packets until the network controller signals to alleviate them. Thus, in the system model, we implicitly also work with $R^+$ instead of $R$.

\section{Simulation}

\begin{figure}[htbp]
				\centering
   				\psfrag{gg}[tl][tl]{\begin{minipage}{2.8cm} \begin{flushleft} channel behavior due to disturbance pattern \end{flushleft}\end{minipage}}
   				\includegraphics[width=0.90\linewidth]{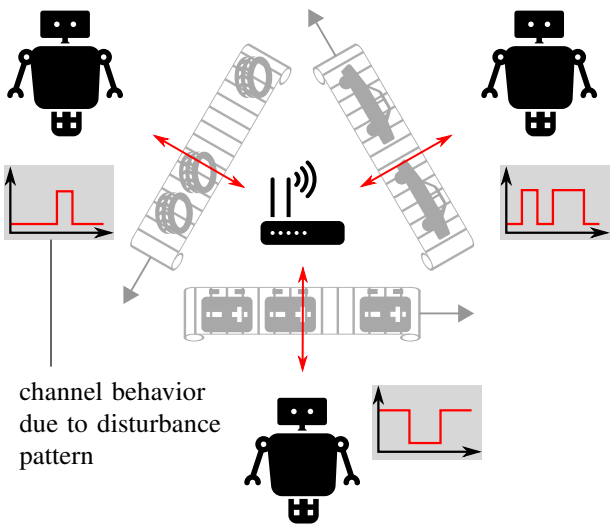}
                \caption{Scenario used for simulations}
                \label{fig::factory}
\end{figure}

For numerical results, we compare the well known MW policy with our introduced predictive network control policy (PNC). We use a scenario in which three robots communicate via wireless connection as depicted in Fig. \ref{fig::factory} and all communication is routed through a central router. Disturbances in the communication are caused by periodic environmental effects, e.g. moving objects in a factory building.

For their work tasks, the robots need to exchange data. We assume that at time $t=0$, each robot needs to send its data (modeled as one packet) to the other two and has signaled this need to the router. The router then assigns communication resources to the robots. Specifically, we assume that in each time slot, \textit{either} one, and only one robot may send its packet to the router (interference property) $\textit{or}$ the router may send a single packet to all robots at once (broadcast property). Note that we imply, that signaling between the agents is instantaneous compared to the transmission of the packets, containing the actual data. This seems to be a reasonable assumption when working with RMPCs, since exchanged data consists of entire trajectories of their internal states.

The performance of PNC depends highly on the periodic disturbances, that dictate the transmission success probabilities $p^i_t$. For this reason we simulate over many randomly selected disturbance patterns and then average the results. We assume that any disturbance pattern has a period of $k$ steps and evolves deterministic in time so that it can be represented (in terms of transmission success probabilities) by a discrete time markov chain with binary transition matrix and a set $\mathcal{M}$ of probability matrices $M_i$, holding the $p^i_t$.
Furthermore we specify, that in each step the transmission success probability is either high $\hat{p}$ or low $\check{p}$ resulting in $2^k-1$ different patterns (we do not consider the unique pattern, only consisting of $\check{p}$). Finally, in order to avoid non-unique solutions to the optimization, we slightly vary $\hat{p}$ for each pattern, once this pattern is selected for simulation, by using a randomly drawn value from the uniform distribution $\left[ \hat{p} \pm 0.01 \hat{p} \right]$ instead of the value $\hat{p}$ itself; the same goes for $\check{p}$.

A \textit{single} simulation run follows the system evolution for $N$ time steps while the PNC policy uses a prediction horizon of $H$. 
We accumulate simulation runs via two loops. The first one repeats over $x$ randomly chosen cases (without repetition). A case is defined as an assignment of patterns to the links of the three robots, resulting in $\left(2^k-1\right)^3$ different cases.
In a second loop (having a fixed case) we simulate over different initializations of the routing matrix (equivalent to initializations of $\mathbb{B} \left[ M_{\sigma_t} \right]$ per link per time slot). We repeat this inner loop $y$ times, resulting in $x \cdot y$ simulation runs.

\subsection{Detailed description of a specific case}

We first demonstrate the general disadvantage of MW on a specific case (i.e. every link has a fixed pattern assigned). We chose the following parameters:

\begin{table}[h]
\centering
\caption{Simulation parameters (specific case)}
\begin{tabular}{ |c|c|c|c|c|c|c|c|c| }
  \hline
  $k$ & $K$ & $\hat{p}$ & $\check{p}$ & $1-\tau$ & $N$ & $H$ & $x$ & $y$ \\
  \hline
  3 & 343 & 100\% & 0\% & 90\% & 20 & 4 & 1 & 1 \\
  \hline
\end{tabular}
%\caption{Simulation Parameters of specific Case}
%\label{fig::simu_parameters}
\end{table}

\begin{figure}
    \centering
    %\hspace{0.6cm}
    \psfrag{R1}[cr][cr]{Robot 1}
    \psfrag{R2}[cr][cr]{Robot 2}
    \psfrag{R3}[cr][cr]{Robot 3}
    \psfrag{ofd}[cl][cl]{optimal first activation}
    \psfrag{npc}[l][l]{
                        \begin{minipage}{2.5cm}
                        \begin{flushleft}
                            no resources available
                        \end{flushleft}
                        \end{minipage}
                        }
    \psfrag{pc}[cl][cl]{
                        \begin{minipage}{2.5cm}
                        \begin{flushleft}
                            available resource block
                        \end{flushleft}
                        \end{minipage}
                        }
    \psfrag{t}[lc][lc]{time}
    \psfrag{period}[cc][cc]{period}
    \psfrag{p11}[c][c]{$p_0^1$}
    \psfrag{p12}[c][c]{$p_1^1$}
    \psfrag{p13}[c][b]{$\dots$}
    \psfrag{p21}[c][c]{$p_0^2$}
    \psfrag{p22}[c][b]{$\dots$}
    \psfrag{p31}[c][c]{$p_0^3$}
    \psfrag{p32}[c][b]{$\dots$}
    \includegraphics[width=0.7\linewidth]{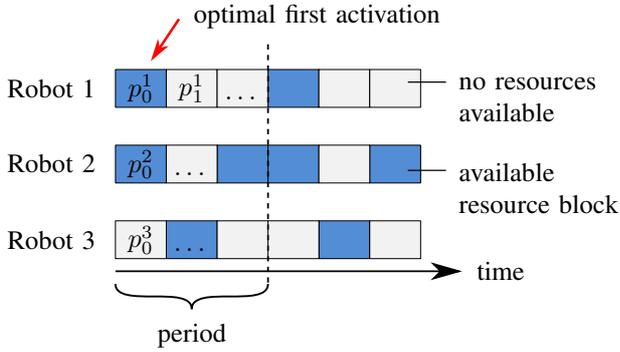}
    \caption{Pattern assignment for specific case}
    \label{fig::sp_resources}
\end{figure}

The patterns are illustrated in Fig. \ref{fig::sp_resources}. The blue colors indicate time slots, in which
$p^i_t = \hat{p} = 100\%$, grey indicates that $p^i_t = \check{p} = 0$. Robot 1 can only communicate once per period; Robot 2 twice. In the described setup, each robot has to send its packet over its link to the router (disjunct actions), before the router can possibly broadcast the packet, using two resource blocks at once. Hence, in the very first time slot, in order to minimize overall delay, it is always optimal to let Robot 1 send its packet to the router, since Robot 2 has the uncontested third time slot to do so and Robot 3 can only communicate in orthogonal time slots. Using PNC, this is indeed always the first action taken.

However, using MW, the decision which Robot (1 or 2) gets to communicate in the very first time slot only depends on the transmission success probabilities $p^1_0$ and $p^2_0$. In practice, if $p^2_0$ is just slightly higher than $p^1_0$, MW will allocate the this first time slot to Robot 2, hence resulting in a sup-optimal control of the network.

\begin{figure}
\begin{tikzpicture}
\begin{axis}[
    height=5cm,
    width=\axisdefaultwidth,
    xlabel={Sequence in which robots fulfill 
    $\hat{p}^{\alpha} > \hat{p}^{\beta} > \hat{p}^{\delta}$},
    ylabel={Ratio of accumulated delay: $\frac{\text{PNC}}{\text{MW}}$},
    xmin=0.5, xmax=6.5,
    ymin=0.71, ymax=0.88,
    xticklabels={
        123 , 132 , 213 , 231 , 312 , 321},
    xtick={1,2,3,4,5,6},
    ytick={ 0.8333 , 0.7576 },
    legend pos=north west,
    ymajorgrids=true,
    grid style=dashed,
]
 
\addplot[
    only marks,
    mark size=2pt,
    color=blue,
    %mark=., %square,
    ]
    coordinates {
    (1,0.8333)(2,0.8333)(3,0.7576)(4,0.7576)(5,0.8333)(6,0.7576)
    };
    %\legend{CuSO$_4\cdot$5H$_2$O}

\end{axis}
\end{tikzpicture}
\caption{Simulation results (specific case)}
\label{fig::sim_results_spec_case}
\myfigurespace
\end{figure}
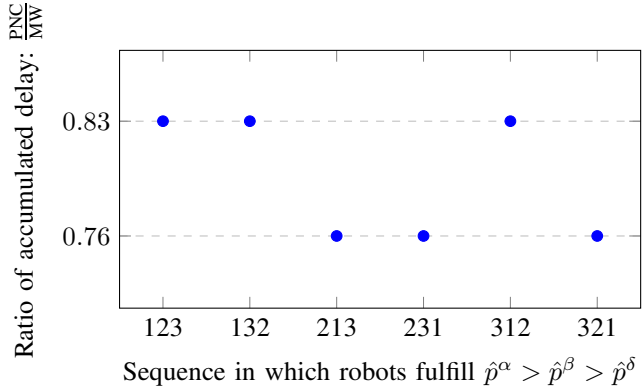

To showcase this, we simulate over all possible slight variations in $\hat{p}$. The behavior of MW might differ, depending on how the robots $1,2,3$ must be mapped to the indices $\alpha, \beta, \delta$ in order to fulfill the inequality $\hat{p}^{\alpha} > \hat{p}^{\beta} > \hat{p}^{\delta}$. There are six different mappings that do that, corresponding to the symmetry group $S_3$. Fig. \ref{fig::sim_results_spec_case} shows simulation results for all six possibilities. MW and PNC are compared by accumulating (over all robot-queues) their respective delays and taking the quotient. For this specfic case, PNC reduces overall delay by about 17 to 24 percent compared to MW, depending on the transmission success probabilities.

% \begin{figure}[htbp]
% 				\centering
%   				\psfrag{p1}[c][c]{17\%}
%   				\psfrag{p2}[c][c]{17\%}
%   				\psfrag{p3}[c][c]{24\%}
%   				\psfrag{p4}[c][c]{24\%}
%   				\psfrag{p5}[c][c]{17\%}
%   				\psfrag{p6}[c][c]{24\%}
%   				\psfrag{MW}[lb][lb]{MW}
%   				\psfrag{PNC}[lb][lb]{PNC}
%   				\psfrag{cases}[c][c]{slightly different $\hat{p}$ for each link}
%   				\psfrag{adt}[c][c]{accumulated delays}
%   				\includegraphics[width=0.8\linewidth]{special_case.eps}
%   				\vspace{2mm}
%                 \caption{Scenario for simulation}
%                 \label{fig::sc_results}
% \end{figure}

\subsection{General simulation results}

Next, we present results for from extended simulation (Monte Carlo simulation) over several cases.

\begin{table}[h]
\centering
\caption{Simulation parameters (Monte Carlo)}
\begin{tabular}{ |c|c|c|c|c|c|c|c|c| }
  \hline
  $k$ & $K$ & $\hat{p}$ & $\check{p}$ & $1-\tau$ &  $N$ & $H$ & $x$ & $y$ \\
  \hline
  3 & 343 & \dots & \dots & \dots & 40 & 4 & 10 & 200 \\
  \hline
\end{tabular}
%\caption{Simulation Parameters of Monte Carlo Simulation}
%\label{fig::simu_parameters_2}
\end{table}

\begin{figure}
\begin{tikzpicture}
\begin{axis}[
    height=6cm,
    width=\axisdefaultwidth,
    xlabel={Global low transmission success probability $\check{p}$},
    ylabel={Ratio of accumulated delay: $\frac{\text{PNC}}{\text{MW}}$},
    xmin=-0.05, xmax=0.55,
    ymin=0.88, ymax=1.04,
    xtick={ 0 , 0.1 , 0.2 , 0.3 , 0.4 , 0.5 },
    ytick={ 0.899 , 0.925 , 0.946 , 0.98 , 1.022},
    legend pos=north west,
    ymajorgrids=true,
    grid style=dashed,
]
\addplot[
only marks,
    color=blue,
    mark size=2pt,%mark=o, %square,
    ]
    coordinates {
    (0,0.899)(0.1,0.925)(0.2,0.946)(0.3,0.977)(0.4,0.984)(0.5,1.022)
    };
    %\legend{CuSO$_4\cdot$5H$_2$O}
\end{axis}
\end{tikzpicture}
\caption{Simulation results (Monte Carlo)}
\label{fig::monte_carlo}
\myfigurespace
\end{figure}
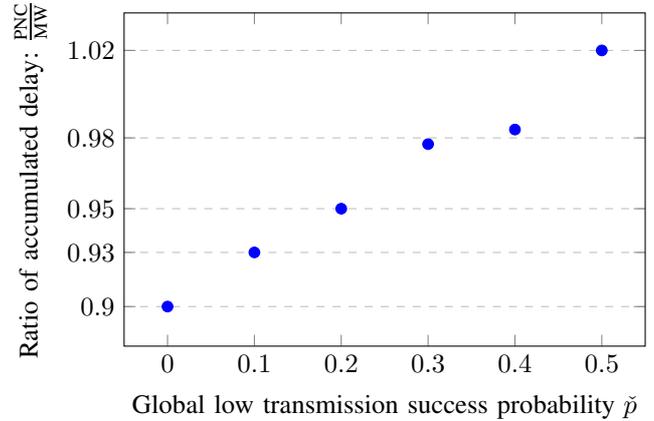

We simulate for different $\check{p}$, where we adjust $\hat{p}$ according to $\hat{p} + \check{p} = 1$; Fig. \ref{fig::monte_carlo} holds the results. Note that we also adjusted the threshold $\tau$ when using a different $\check{p}$, so that PNC always deems $\hat{p}$ reliable, i.e. $\tau > 1 - \hat{p}$. Not adjusting $\tau$ leads to a distinct drop in performance of the PNC policy, since the short horizon of $H=3$ does not suffice to schedule most of the transmission in a reliable way. In other words, a high reliability requirement (in comparison to the available transmission success probabilities), has to be accompanied with a far enough horizon to enable the algorithm to reliable schedule in its prediction model.

The simulations show, that we can expect an average reduction in accumulated delays of about 10\%, if transmission success probabilities (channel states) jump between $1$ (superb) and $0$ (not available at all). The closer $\check{p}$ and $\hat{p}$ get, the more this pleasant reduction diminishes. In the instance, that $\check{p} = \hat{p} = 0.5$, MW even exceeds the performance of PNC. This result is due to the fact, that in this instance, future predictions are the least helpful (there is no time dependent pattern to take advantage of and the one step optimal control becomes the general optimal control).

Note that the resulting quotient of a single simulation can differ heavily (0.5 to 1.5) from the obtained averaged performance quotients (0.9 \dots 1.2). Also, the discussion above does not take into account, that we additionally yield individual forecasts of delays. One should keep in mind, that the reduction of accumulated delay is only one benefit of the PNC algorithm. And finally, the simulated scenario resembles a bursty stimulation of the network. The disadvantages of MW become less stringent, once the bursty traffic transitions into a steady state traffic, because then, MW can use the length of individual queues to obtain information on good and bad paths through the network.

\subsection{Time consumption}

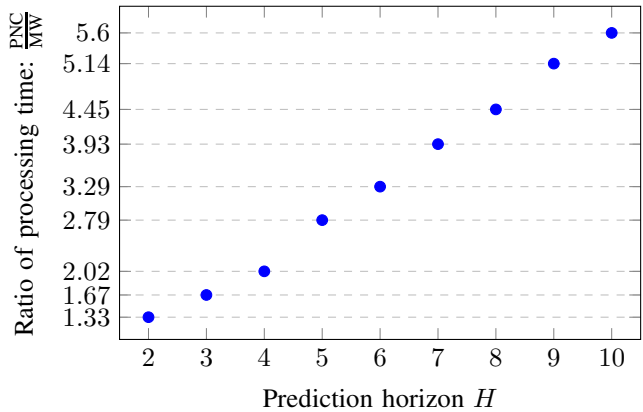
\begin{figure}
\begin{tikzpicture}
\begin{axis}[
    height=6cm,
    width=\axisdefaultwidth,
    xlabel={Prediction horizon $H$},
    ylabel={Ratio of processing time: $\frac{\text{PNC}}{\text{MW}}$},
    xmin=1.5, xmax=10.5,
    ymin=1, ymax=6,
    xtick={  2,3,4,5,6,7,8,9,10 },
    ytick={ 1.333947949 , 1.667348364 , 2.022169807 , 2.790618529 , 3.292964822 , 3.931115132 , 4.452173822 , 5.139167967 , 5.600196264 },
    legend pos=north west,
    ymajorgrids=true,
    grid style=dashed,
]
\addplot[
only marks,
    color=blue,
    mark size=2pt,%mark=o, %square,
    ]
    coordinates { (2,1.333947949) 
    (3,1.667348364)
    (4,2.022169807)
    (5, 2.790618529)
    (6, 3.292964822)
    (7, 3.931115132)
    (8, 4.452173822)
    (9, 5.139167967)
    (10, 5.600196264) };
    %\legend{CuSO$_4\cdot$5H$_2$O}
\end{axis}
\end{tikzpicture}
\caption{Simulation results (time consumption over horizon)}
\label{fig::time_consumption}
\myfigurespace
\end{figure}

By applying the binary linear optimization over the horizon $H$, the minimization problem in PNC has to be solved for $m\cdot H$ unknown binary values. In comparison, MW does only solve for $m$ unknown binary values, since in each step it solves
\begin{equation}
    \min_{u} q^T R u
\end{equation}
where $q$ and $R$ are current queue vector and current routing matrix. Though one would intuitively suspect an exponential growth (with $H$) in time needed for deriving at an optimal solution, at least for scenario presented here, simulations suggest a linear growth as shown in Fig. \ref{fig::time_consumption}. The used parameters are captured in Table \ref{tab::simu_parameters_3}, where we chose $N=7$ to ensure that there are always packets still to be transmitted. If all packets are transmitted, then the consecutive optimization is trivial which would in turn compromises the simulation results.

\begin{table}[h]
\centering
\caption{Simulation parameters (time consumption)}
\begin{tabular}{ |c|c|c|c|c|c|c|c|c|c| }
  \hline
  $k$ & $K$ & $\hat{p}$ & $\check{p}$ & $1-\tau$ & $N$ & $H$ & $x$ & $y$ \\
  \hline
  3 & 343 & 70\% & 30\% & 68\% & 7 & \dots & 10 & 100 \\
  \hline
\end{tabular}
%\caption{Simulation Parameters of Monte Carlo Simulation}
\label{tab::simu_parameters_3}
\end{table}

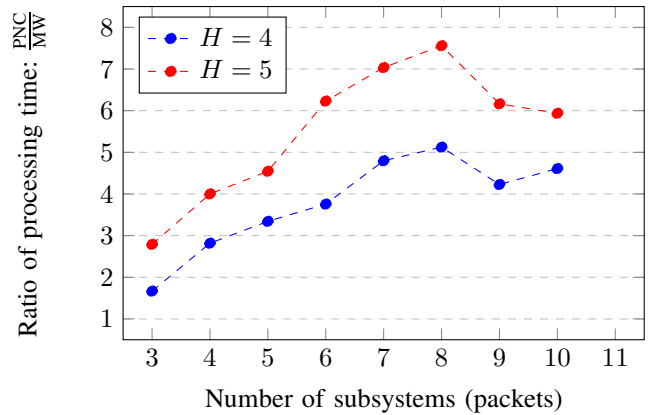
\begin{figure}
\begin{tikzpicture}
\begin{axis}[
    height=6cm,
    width=\axisdefaultwidth,
    xlabel={Number of subsystems (packets)},
    ylabel={Ratio of processing time: $\frac{\text{PNC}}{\text{MW}}$},
    xmin=2.5, xmax=11.5,
    ymin=0.5, ymax=8.5,
    xtick={ 3,4,5,6,7,8,9,10,11 },
    ytick={ 1 , 2 , 3 , 4 , 5 , 6 , 7 , 8 },
    legend pos=north west,
    ymajorgrids=true,
    grid style=dashed,
]
 \addplot[
%only marks,
    color=blue,
    mark=* ,
    mark size=2pt,%mark=o, %square,
    dashed,
    ]
    coordinates { 
(3, 1.667348364)
(4, 2.817641342)
(5, 3.342565206)
(6, 3.75682138)
(7, 4.797910828)
(8, 5.125028812)
(9, 4.227666133)
(10, 4.609873457)
 };
 \addplot[
%only marks,
    color=red,
    mark=* ,
    mark size=2pt,%mark=o, %square,
    dashed,
    ]
    coordinates { 
(3, 2.790618529)
(4, 4.001513475)
(5, 4.545074863)
(6, 6.23085932)
(7, 7.035096623)
(8, 7.559522982)
(9, 6.164559214)
(10, 5.936891772)
 };
%2.790618529 , 4.001513475 , 4.545074863 , 6.23085932 , 7.035096623 , 7.559522982 , 6.164559214 , 5.936891772
    \legend{$H=4$,$H=5$}
\end{axis}
\end{tikzpicture}
\caption{Simulation results (time consumption over subsystems)}
\label{fig::time_consumption_noss}
\end{figure}

Finally, we also try to investigate how time consumption scales with the number of subsystems in the prediction model, i.e. with the number of packets to be transmitted simultaneously. We use again the parameter set from Table \ref{tab::simu_parameters_3} but vary the number of packets, for which transfer is requested in the very first time slot; the results are shown in Fig. \ref{fig::time_consumption_noss}. The casual decrease in time consumption with growing number of packets might be a consequence of the utilized optimizer (gurobi) applying a branch-and-bound procedure to solve the minimization. This remains to be analyzed. Nevertheless, the results once more suggest a linear growth in time consumption with increasing number of packets.

\myfigurespace
\section{Conclusion}

We provided a proof of concept for a new network control policy, which is predictive in nature (based on MPC paradigms), and does provide reliable forecast of delay times of single data packets. The applied optimization problem is linear and thus quite feasible to implement. The numerical results show a clear advantage of our approach in comparison to MW when it comes to pure routing and scheduling decisions. However these advantages are leveraged with an increased utilization of computational resources, the dimension of which we could identify.

%If these disadvantages outweigh the advantages depends on whether the predictive nature of the PNC policy can be used for performance gains (which in turn depends on the network features) and on the size of the data, since high data sizes mean longer transmission times and therefore, relatively, less computational time, making it less costly to use PNC.

\section*{Acknowledgment}

This research is partially supported by the EU H2020-ICT2016-2 project ONE5G and the DFG Priority Programme 1914 Cyber-Physical Networking (CPN). The views expressed in this paper are those of the authors and do not necessarily represent the project views.

%\section*{References}

% Please number citations consecutively within brackets \cite{b1}. The 
% sentence punctuation follows the bracket \cite{b2}. Refer simply to the reference 
% number, as in \cite{b3}---do not use ``Ref. \cite{b3}'' or ``reference \cite{b3}'' except at 
% the beginning of a sentence: ``Reference \cite{b3} was the first $\ldots$''

% Number footnotes separately in superscripts. Place the actual footnote at 
% the bottom of the column in which it was cited. Do not put footnotes in the 
% abstract or reference list. Use letters for table footnotes.

% Unless there are six authors or more give all authors' names; do not use 
% ``et al.''. Papers that have not been published, even if they have been 
% submitted for publication, should be cited as ``unpublished'' \cite{b4}. Papers 
% that have been accepted for publication should be cited as ``in press'' \cite{b5}. 
% Capitalize only the first word in a paper title, except for proper nouns and 
% element symbols.

% For papers published in translation journals, please give the English 
% citation first, followed by the original foreign-language citation \cite{b6}.

\bibliography{globe_com_paper}

\begin{thebibliography}{1}
\providecommand{\url}[1]{#1}
\csname url@samestyle\endcsname
\providecommand{\newblock}{\relax}
\providecommand{\bibinfo}[2]{#2}
\providecommand{\BIBentrySTDinterwordspacing}{\spaceskip=0pt\relax}
\providecommand{\BIBentryALTinterwordstretchfactor}{4}
\providecommand{\BIBentryALTinterwordspacing}{\spaceskip=\fontdimen2\font plus
\BIBentryALTinterwordstretchfactor\fontdimen3\font minus
  \fontdimen4\font\relax}
\providecommand{\BIBforeignlanguage}[2]{{%
\expandafter\ifx\csname l@#1\endcsname\relax
\typeout{** WARNING: IEEEtran.bst: No hyphenation pattern has been}%
\typeout{** loaded for the language `#1'. Using the pattern for}%
\typeout{** the default language instead.}%
\else
\language=\csname l@#1\endcsname
\fi
#2}}
\providecommand{\BIBdecl}{\relax}
\BIBdecl

\bibitem{hahn2018}
J.~Hahn, R.~Schoeffauer, G.~Wunder, and O.~Stursberg, ``Distributed mpc with
  prediction of time-varying communication delay,'' in \emph{IFAC Workshop on
  Distributed Estimation and Control in Networked Systems (NecSys)}, 2018.

\bibitem{tassiulas1992}
L.~Tassiulas and A.~Ephremides, ``Stability properties of constrained queueing
  systems and scheduling policies for maximum throughput in multihop radio
  networks,'' \emph{IEEE Transactions on Automatic Control}, 1992.

\bibitem{meyn2007}
S.~Meyn, \emph{Control Techniques for Complex Networks}.\hskip 1em plus 0.5em
  minus 0.4em\relax Cambridge University Press, 2007.

\bibitem{kasparick2017}
M.~Kasparick and G.~Wunder, ``Stable wireless network control under service
  constraints,'' \emph{IEEE Transactions on Control of Network Systems}, 2017.

\bibitem{zhou2007}
C.~Zhou and G.~Wunder, ``Throughput-optimal scheduling with low average delay
  for cellular broadcast systems,'' in \emph{IEEE GlobeCom 2007 - IEEE Global
  Telecommunications Conference}, 2007.

\bibitem{frangioni2017}
A.~Frangioni, L.~Galli, and G.~Stea, ``Delay-constrained routing problems:
  Accurate scheduling models and admission control,'' \emph{Computers and
  Operations Research}, 2017.

\bibitem{lee2009}
J.~Lee and N.~Jindal, ``Delay constrained scheduling over fading channels:
  Optimal policies for monomial energy-cost functions,'' \emph{IEEE
  International Conference on Communications}, 2009.

\bibitem{schoeffauer2018}
R.~Schoeffauer and G.~Wunder, ``Predictive network control and throughput
  sub-optimality of max weight,'' in \emph{2018 European Conference on Networks
  and Communications (EuCNC)}, June 2018, pp. 1--6.

\bibitem{munapo2016}
E.~Munapo, ``Solving the binary linear programming model in polynomial time,''
  \emph{American Journal of Operations Research}, 2016.

\end{thebibliography}
\bibliographystyle{IEEEtran}

\end{document}